\documentclass[12pt]{article}

\usepackage{jinstpub} 
                     
\usepackage{hyperref}

\title{\boldmath Signal Processing to Reduce Dark Noise Impact in Precision Timing}

\author[a]{Sebastian~White}

\emailAdd{sebastian.white@cern.ch}

\affiliation[a]{ University of Virginia \\Charlottesville, Virginia, USA}

\abstract{ 
	We introduce a technique to mitigate the effects of low frequency noise on precision timing.
The example of Dark Count Noise Rate (DCR) in Silicon Photomultipliers (SiPMs) is emphasized.
This technique exploits the correlation between time shifts on
the leading edge of a signal and the residual slope of the baseline (due to noise) which remains
after baseline subtraction.

	In fast timing applications (such as for Time-of-flight particle ID) the signal arrival time is typically captured on the signal leading edge. 
	The signal risetime is often fixed by the physics of the sensor and input circuit. Then accurate pulse timing can be achieved by correcting a leading edge threshold time (depending on a slope proportional to both the Amplitude and the risetime) to a ``constant fraction" time. 	
	This compensation for time walk due to amplitude fluctuations breaks down once we introduce interference from low frequency noise on the leading edge. In this paper we demonstrate that an additional measurement of the slope at threshold can be used to correct for this noise jitter. }
	
\keywords{ Timing detectors, SiPM, Signal Processing}

\begin{document}
\maketitle
\flushbottom
\section{Introduction}
\label{sec:Introduction}

In the 1990's the Large Hadron Collider (LHC) community turned to designing experiments capable of a $\sim2$ order of magnitude increase in interaction rates over the SppS and Tevatron colliders \cite{Shiltsev}. This enabled covering the full allowed Higgs Mass range- supplementing the beam energy increase available at the pre-existing LHC ring with unprecedented luminosity. In addition to the demands on trigger sophistication and sensor robustness that come with higher luminosity a notion that enters design for the High Luminosity LHC (HL-LHC) and future colliders is "End of Life Performance".  

	Rather than designing signal processing with a sole focus on optimal performance under Day-1 conditions,
we introduce also tools which address the performance degradation of sensors after extended exposure to radiation. In this paper we introduce a technique that mitigates (timing) performance degradation due to  low frequency noise in radiation damaged SiPMs . 

\subsection{New Aspects of Silicon Radiation Damage}

	The widespread use of Silicon sensors in collider applications has been enabled by the study of displacement damage in
Silicon diodes over the past 20 years or so \cite{Moll}.  In many cases integrated doses reach as high as $10^{15}$ neutron equivalent per cm$^2 $ (neq/cm$^2$). The main consequence has been increased leakage current with well characterized dependence on temperature and partial recovery by annealing.

	With the growing use of Silicon devices with internal gain (Avalance Diodes and Geiger mode Avalanche Photodiodes- aka SiPMs), other aspects of radiation damage received attention. At fluences above  $\sim10^{14}$ neq/cm$^2$, which is often the case in current and planned experiments, the charge multiplication in Silicon (Gain) is degraded (to be compensated with higher operating voltage).

	In SiPM applications exposed to these or higher fluences (ie ATLAS and CMS HL-LHC upgrades-  such as the CMS High Granularity Calorimeter and Barrel Timing Layer (BTL )) \cite{TDR} the increased leakage current is converted to a stream of pulses ("Dark Counts") consisting of individual charges amplified by Geiger mode operation to typically $\sim 10^5$ e$^-$. The individual pulses are typically in the form of a spike with exponential decay constant, $\lambda$, according to the RC parameters of the SiPM. This pulse shape can be observed directly when the Dark Count rate is low. However for the radiation exposures relevant to these LHC upgrade examples, the Dark Count Rate (DCR) reaches 10s of GHz and what is observed is a chaotic, mostly low frequency, ripple which we will refer to as "Dark Count Noise". An example is shown in figure 5 below.

	Analysis of the frequency spectrum of this sort of noise can be traced back to a 1926 paper of Schottky \cite{Schottky} where, starting from the Fourier transform of a single pulse:

\begin{equation}
F(\omega )=\int_{-\infty}^{\infty}V(t) e^{-i\omega t}  \,dt\ = A\int_{0}^{\infty} e^{-i( \lambda+\omega ) t}  \,dt\ = \frac{A}{\lambda+i\omega}
\end{equation}
he derives a power spectrum, for a train of such pulses:

\begin{equation}
S(\omega)=\lim_{T\to\infty} \frac{1}{T}\langle \mid F(\omega)\mid^2 \rangle =\frac{A^2}{\lambda^2+\omega^2}\lim_{T\to\infty} \frac{1}{T}\langle \mid \sum_{k}^{} e^{i\omega t_k}\mid ^2\rangle
=\frac{n\times A^2}{\omega^2+\lambda^2}
\end{equation}

proportional to the pulse rate, n. The triangle brackets denote an average of trials.  Above a frequency, controlled by $\lambda$ the Power spectrum follows $S[f]\sim \frac{1}{f^2}$ yielding the behavior of the Discrete Fourier Spectrum (ie $S^{\frac{1}{2}}$) that we will see in Figure 4.\footnote[1]{ In the literature this analysis is sometimes extended to the case where a range of values for $\lambda_i ...\lambda_n$ are involved. We then obtain a Power Spectrum with a 1/f dependence, for which there is an extensive literature.}

 	While there are
signal processing approaches which would minimize the effect on energy measurement \cite{Radeka}, these do not apply to the very significant impact that this low frequency noise will have on precision time measurements in the LHC upgrades, as discussed below.

As this paper is concerned with timing detectors for the LHC and introduces several common techniques (such as "Amplitude Walk Correction") the reader may find in \cite{Jerry} a general overview of the current technology and techniques for optimizing timing in the picosecond regime. The motivation for timing upgrades in ATLAS and CMS can be found in \cite{intro,Uli,seb} as well as \cite{TDR}.

The technique we will demonstrate here first became evident with LYSO/SiPM prototype sensors for the CMS Barrel Timing Layer exposed to a 120 GeV proton beam. However the approach of this paper will be
to demonstrate the tool for time jitter reduction:
\begin{itemize}
\item{ starting from model data for signal and noise- We establish the case that low frequency noise is amenable to a correction procedure.} 
\item{then introduce the actual noise spectrum from a SiPM with 20 GHz Dark Count Rate}
\item{then use the full laser data for the signal and noise}
\item{ and, lastly, turn to LYSO/SiPM prototype test beam data}
\end{itemize}

\section{Data Sets Used for this Paper}

	For most of the discussion below we refer to a set of laboratory measurements reported earlier \cite{IPRD} employing a 3x3 mm$^2$ SiPM\footnote[2]{Hamamatsu  HPK S12572 - 015 SiPM} with a high bandwidth transimpedance amplifier. Waveforms were recorded using a 1 GHz analog bandwidth, 20 GSa/s digital oscilloscope. The SiPM was illuminated from a PicoQuant Laser head with 470 nm peak emission and a pulse width of 350 picoseconds (rms). Lastly, the induced Dark Count Rate, was generated by concurrently exposing the SiPM to a DC light source. Further details can be found in \cite{IPRD}.

	A similar SiPM model with identical readout electronics and digital scope was used in 2020 at the Fermilab test beam facility in a beam of 120 GeV protons. The detector prototype consisted of a 3x3x57 mm$^3$ bar of Ce doped LYSO  with one SiPM on either end. The beam traversed the bar at a shallow angle (53$^{\circ}$ relative to the normal) corresponding to a typical angle of incidence in the planned HL-LHC upgrade of the CMS Barrel Timing Layer. The average time of SiPMs was compared to a reference MicroChannel plate PMT \cite{Lukas}. Measurements of the time resolution were performed with both unirradiated SiPMs (Dark Count $\sim 0$) and radiation damaged SiPMs  (Dark Count $\sim 13$ GHz) .

	Throughout this paper time measurements are based on a sample of 1-5 k events/waveforms. The "signal arrival time" refers to a measurement of the time the signal crosses threshold at a constant fraction (CF) of the peak amplitude relative to a reference "t$_0$" given by the laser trigger or, when placed in a particle beam,  an MCP-PMT.

	Finally, the "time resolution" measurement is the rms ($\sigma$ ) obtained from fitting those data to a Gaussian (which is always a good description of the data).

\begin{figure}[!htp]
\includegraphics[width=.65\textwidth]{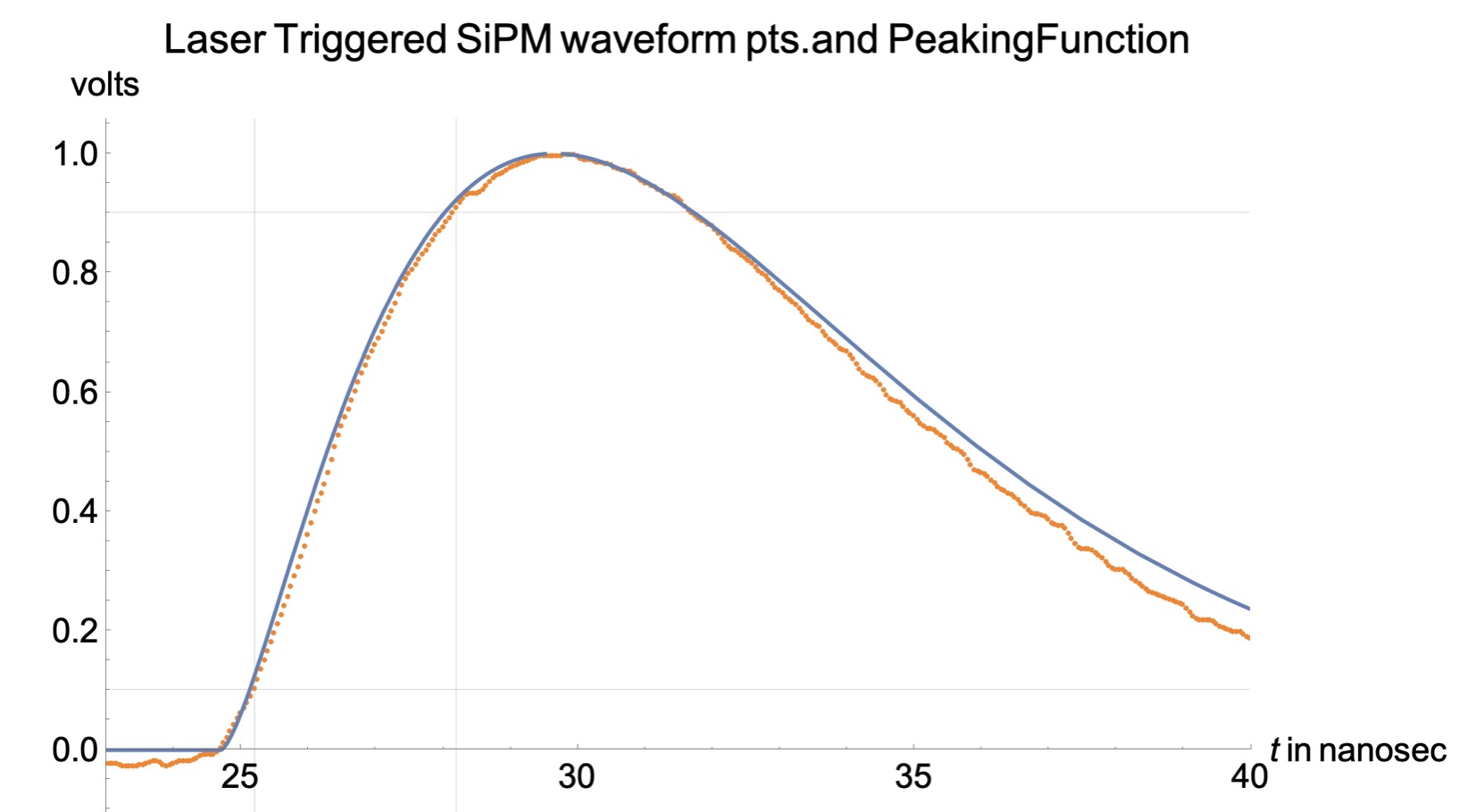}
\includegraphics[width=.25\textwidth]{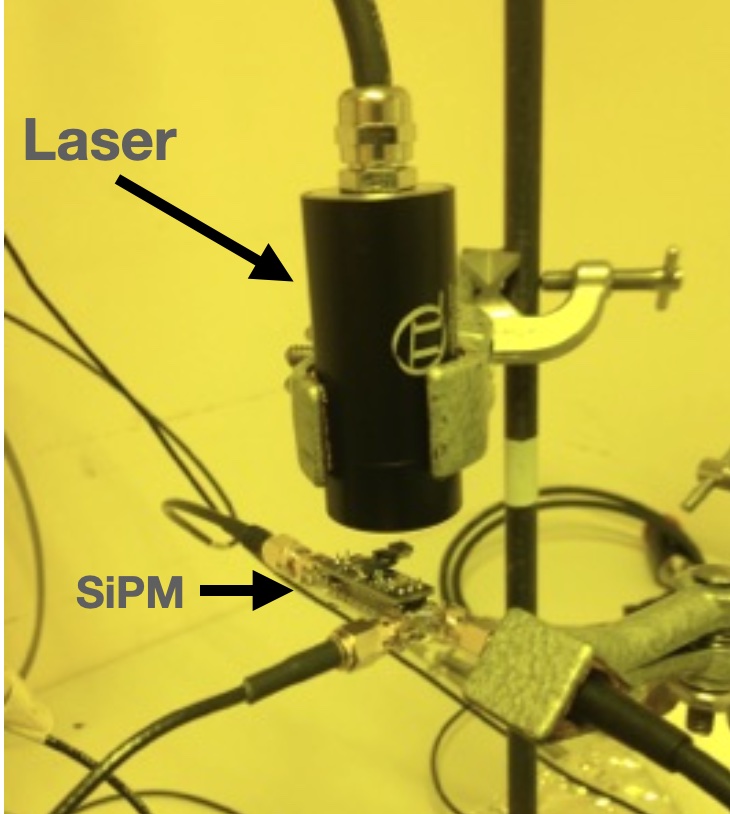}
\caption{For the initial analysis with a model signal and both the simple noise model and actual noise we use a template for the laser signal (solid, blue in left figure). In this figure a typical waveform is compared with the model ("Peaking Function" defined in the text). A laser with peak emission at 470 nm directly illuminated the SiPM (right).}
\label{fig:wavefor}       
\end{figure}

\subsection{ Fixed Template for Laser signal and a simple noise model}

	Starting with the Laser data, we will use data sets with  $<N_{pe}> \sim$ 148 or 390 (measured as described in \cite{IPRD}) is the mean signal delivered by the laser head.
Of course, when examined in detail, the signal itself can contribute to time jitter in the measurement (as will be employed throughout) of leading edge timing on an "Amplitude corrected" signal. For example, fluctuations at the leading edge arising from photostatistics contribute. Also amplitude fluctuations are corrected for by renormalizing to 1V peak amplitude. So, effectively, for the rest of the discussion a signal/threshold level really refers to a Constant Fraction of the Peak.


	Since we first want to demonstrate how low frequency noise impacts timing measurements and possible tools to reduce the impact on jitter, we start with a model. As shown in Figure 1, a typical laser signal is very well approximated by a so-called "Peaking Function" with the right choice of parameters:

\begin{equation}
F[t]=A\times (\frac{t-t_0}{\tau})^n\times Exp[-\frac{(t-t_0)}{\tau}]  
\end{equation}

where n=1.5, $\tau=3.3$ nanosec, and for (t-t$_0$) positive. In Figure 1 the points are a typical digitized waveform and the solid curve is the above function.


	How will low frequency ripple due to a Dark Count Rate (DCR) impact leading edge timing measurements on this waveform?
Let's start with the simplest case of a fixed frequency sine wave with arbitrary phase as shown in Figure 2.

	A key aspect of signal processing in the presence of high rates and low frequency noise is the use of baseline estimation/subtraction.
This is necessary because otherwise offsets at the start of the signal would make it impossible to make meaningful timing
(or energy) measurements. In this report we will employ two different techniques to accomplish the subtraction since, as we will see, the tool we are
demonstrating for mitigation of jitter from Dark Noise is closely linked to the subtraction.

 	In the first case subtraction is based on direct inspection of the waveform prior to the signal . In the second example we use the more practical technique of "Single Delay Line Shaping" \cite{Knoll}. In the latter, a copy of the signal is delayed by a fixed amount and subtracted from the original, which automatically removes baseline offset. A detailed account using the SDL technique on this data set can be found in \cite{IPRD}.

	In the following we will show that with either technique there is an effective tool to compensate for time jitter
due to this low frequency noise. The missing element is that the usual techniques don't capture the residual \underline{slope} of the baseline noise
even if they correct for the average \underline{level}.

A residual baseline slope results in distortion of the pulse leading edge and consequently degradation of leading edge timing. The subtraction, by resetting the baseline offset (due to noise),
ensures that the value of the noise offset for $t>t_0$ will be locked to the sign of the slope offset.

In this paper we demonstrate that some of this loss in time precision can be recovered if the baseline slope (or, more practically, the slope at timing threshold) can also be
measured.

\begin{figure}
\centering
\centerline{\includegraphics[width=0.95\textwidth]{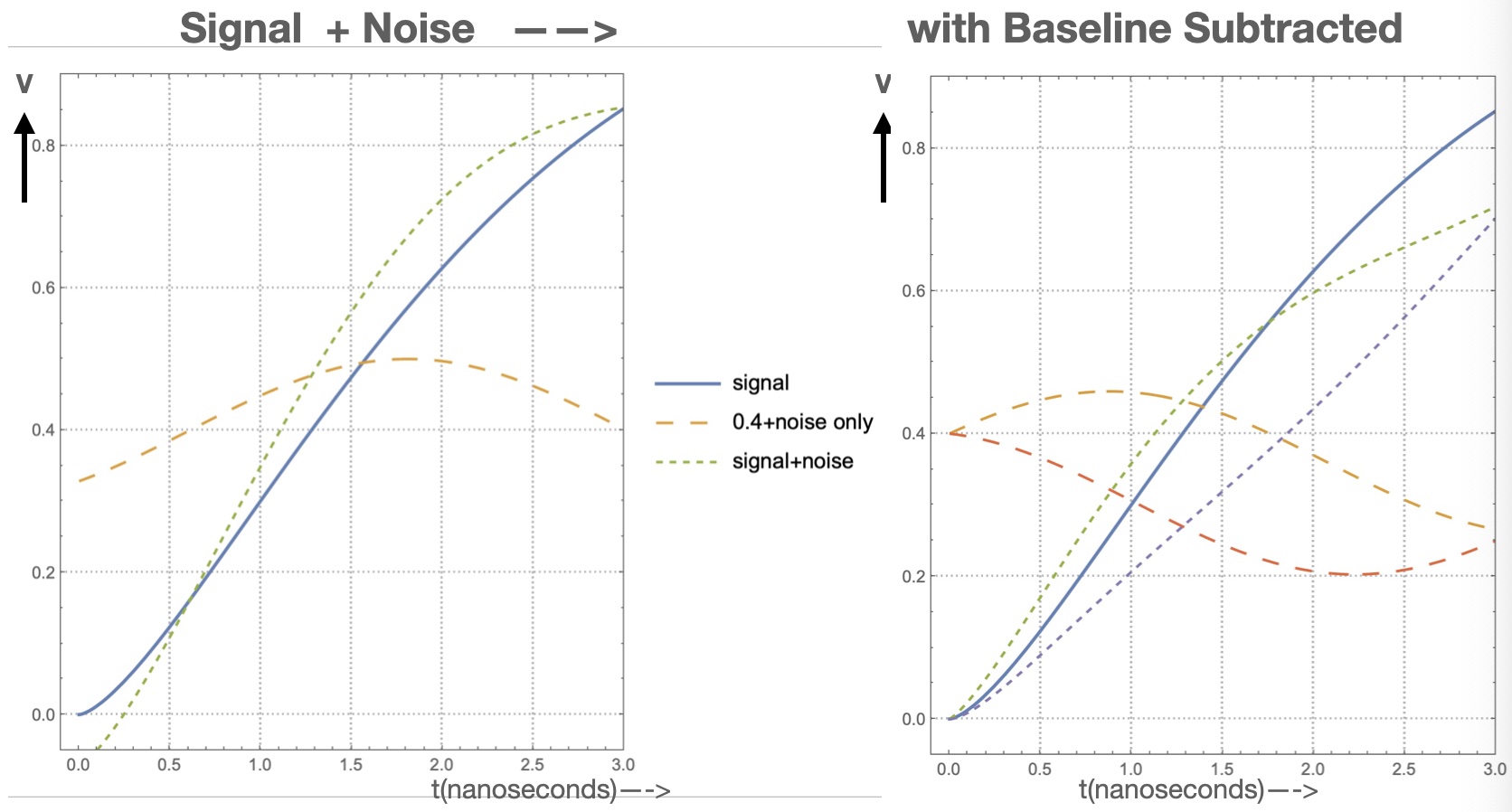}}
\caption{The signal leading edge (expanded from Figure 1) with a sine wave for the noise model added.(The latter is shifted by 0.4V for this illustration). In the left figure we see that for early times (ie 0.3 nanosec) the deviation in 1) slope and 2) amplitude (and, consequently, time at fixed threshold) can occur with arbitrary relative sign. In the example depicted on the left the amplitude offset is negative while the slope offset is positive. The right panel shows that baseline subtraction forces the change in 1) slope and 2) amplitude at some early time to have the same sign. 
In the right panel we consider two arbitrary values for the phase to demonstrate this. Figure 3, below, illustrates the correlation for arbitrary phase. }
\label{fig:modelmodeln}       
\end{figure}

\subsection{ Threshold time correlation with Slope at Threshold}

For the moment, let's consider the case where "signal waveforms" are identical event-by-event and only the noise background fluctuates.
In fact, for our purposes, this is effectively the case so long as the signal \underline{shape} is invariant and we have the tools to correct for amplitude variations 
("Amplitude walk Correction" \cite{Knoll}).

In the following, we use the Peaking Function (eqn. 3.1) as a model for the signal leading edge before turning to actual data. Let's take the signal and noise model shown in Figure 2.
The noise is a simple sine wave with arbitrary phase relative to the signal (Figure 2 left).
If we examine the distorted signal (= signal+noise) at a particular time, t\textsubscript{threshold} , in the leading edge (usually early for timing applications) the amplitude will be shifted by a +'ve or -'ve amount, V\textsubscript{noise}[t\textsubscript{threshold}]. Consequently the time at which a fixed threshold discriminator fires will be shifted by a -'ve or +'ve amount relative to the nominal time.\footnote[2]{In the following we use the term "t\textsubscript{threshold} "  or "t\textsubscript{thr} ", to denote the nominal time at which a fixed (Voltage) threshold would have fired in the unperturbed signal.}

Could a measurement of the noise slope ( in practical use-- measurement of signal+noise slope relative to the nominal signal slope)  at t\textsubscript{threshold} be used to detect and correct for this time shift?

The answer is clearly "no" for the case of Figure 2 left. So long as the phase is arbitrary the time shift and the slope shift could equally well have the same or opposite signs\footnote[3]{ Recalling that "slope" refers to V'[t$_{thr}$] where the input V[t] contains signal, noise or both.}.

However this picture changes completely once we apply baseline subtraction (see Figure 2 right). Now the noise signal at t\textsubscript{threshold} :

\begin{equation}
V_{noise}[t_{thr}]=A\times sin[\omega \times t_{thr}+\phi]
\end{equation}

with baseline correction becomes
\begin{equation}
V_{noise}[t_{thr}]=A\times sin[\omega \times t_{thr}+\phi]-A\times sin[\phi]
\end{equation}

and for small intervals from t=0 to t\textsubscript{threshold}

\begin{equation}
V_{noise}[t_{thr}]\sim V'[t=0]\times t_{thr} \sim  V'[t=t_{thr}]\times t_{thr}
\end{equation}

So $\Delta$ V and $\Delta$ Slope are now, at least initially, correlated, which, for a given Voltage threshold, causes a time shift of

\begin{equation}
\delta t=-\frac{V_{noise}[t_{thr}]}{dV/dt}
\end{equation}

with

\begin{equation}
dV/dt |_{t_{thr}}=K+V'_{noise}[t_{thr}]
\end{equation}

where K is the, roughly constant, slope of the unperturbed signal and typically K$>V'_{noise}$. So substituting from above

\begin{equation}
\delta t=-\frac{V'_{noise}\times t_{thr}}{K+V'_{noise}}
\end{equation}
\begin{equation}
\delta t \sim -V'_{noise}\times \frac{t_{thr}}{K}
\end{equation}

Having this formula in hand we could now develop a timing correction for noise jitter, analogous to the Amplitude walk correction.

\begin{figure}
\centering
\centerline{\includegraphics[width=0.95\textwidth]{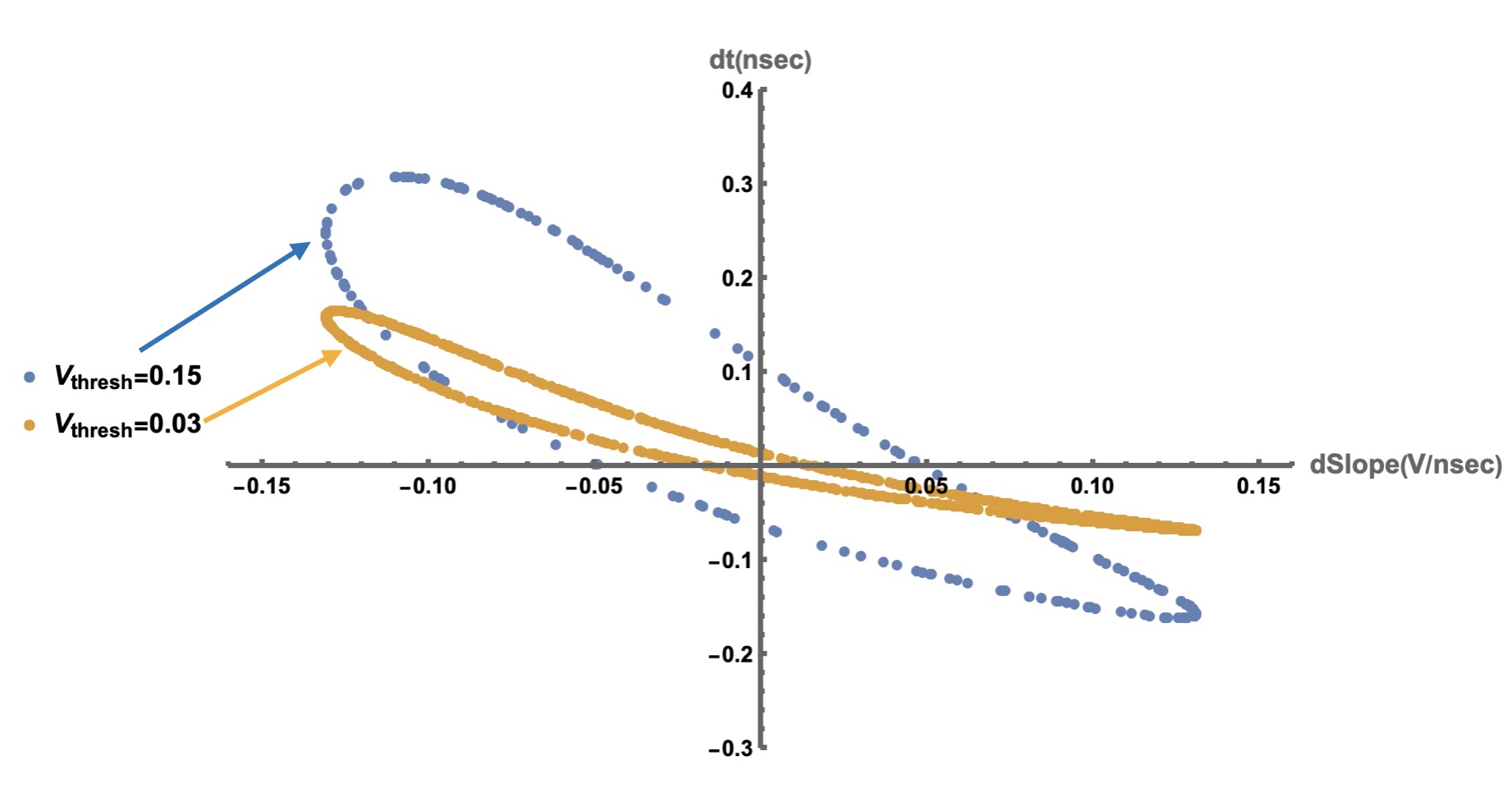}}
\caption{Correlation observed between time deviation and and slope deviation using the simple model shown in Figure 2. For a given noise frequency the correlation follows an orbit with tightest correlation for low threshold values and degrades for higher threshold ($3\%$ and $15\%$ Constant Fraction, respectively).}
\label{fig:wavefor2}       
\end{figure}

	This requires first a subsidiary measurement of V'[t$_{thr}$]  from which we extract the noise part, $V'_{noise}$[t$_{thr}$], by 
subtracting K. In practice, K is obtained empirically as an average slope near a given threshold. For a realistic waveform as well as the models used here, K varies smoothly along the
leading edge. We expect the correction coefficient relating time shift to the measured $V'_{noise}$ to depend somewhat on the threshold setting as will be seen in examples 
below. In fact, these features will be clearly seen in Figure 11.

\subsection{Conclusion from Model Signal and Model Noise}

	Returning to Figure 2, where we take a fixed frequency for the noise term, we can exhibit the limitations of the above approximation that t\textsubscript{threshold} is close to 0.
	
	In Figure 3 we consider two different threshold values for the timing measurement (3\% and 15\% Constant Fraction levels). As we expect, the strongest correlation between
slope and time shift occurs when timing close to t=0. For later times the correlation becomes weaker.

\begin{figure}
\centering
\centerline{\includegraphics[width=0.9\textwidth]{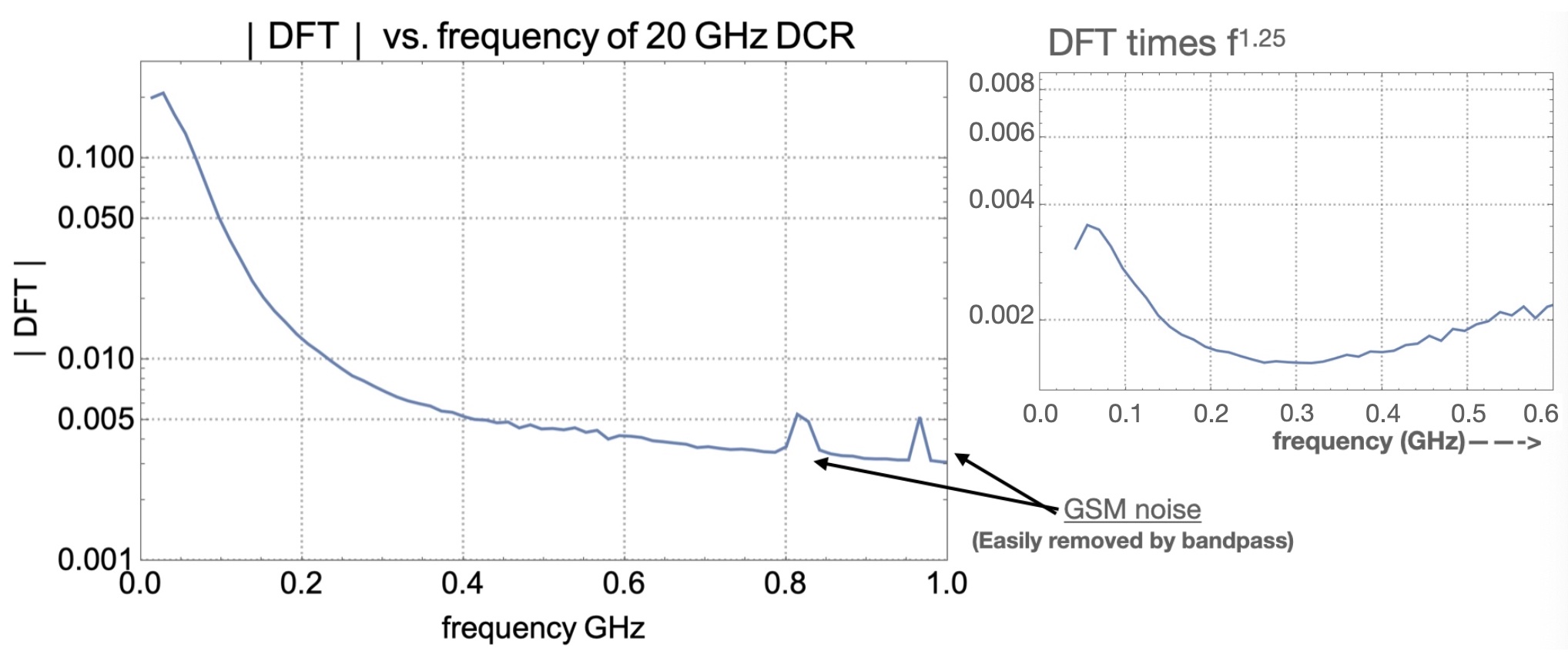}}
\caption{Discrete Fourier transform of the Dark Count generated noise observed in the lab. The right plot shows that the DFT spectrum is characterized by $\sim 1/f$. Note also the presence
of high frequency noise pickup at $\sim 1 $GHz. This high frequency noise is not intrinsic to the system but is an artifact of external pickup which we can turn on and off by selecting the bandpass 
in signal processing. We use this below to demonstrate the importance of proper bandpass matching to the signal. }
\label{fig:wavefor3}       
\end{figure}

	The general conclusion from the above model is that in a system incorporating "baseline subtraction", low frequency noise disrupts leading edge time resolution in a manner for which
additional measurement can provide a correction.

\subsection{Extending to Actual Noise data with Full Frequency Spectrum}

We now turn to a realistic noise frequency spectrum from noise waveforms recorded in a laser lab at CERN \cite{IPRD} with a Hamamatsu 3x3 mm\textsuperscript{2} Silicon Photomultiplier (SiPM) of  \cite{IPRD} and Dark Count Rate of 20 GHz as described above.

What can we say about the frequency spectrum of the resultant dark count noise?

In Figure 4 we show the discrete Fourier Transform of noise traces recorded on a high bandwidth digital scope (BW=1 GHz, Sampling Rate 20 GSa/s). This noise spectrum is very similar
to that of the signal (not shown) and is representative of the 1/f\textsuperscript{2} Power Spectrum of eqn. 1.2, as illustrated in Figure 4 (right).

Given the shape of dark count pulses (fast leading edge and exponential fall-off) SiPM Dark Count pulses are a text book example for Schottky's original derivation of the noise spectrum \cite{Schottky} as repeated in eqn. 1.1 above. There is further discussion of the literature in the Appendix below.

\begin{figure}[!htp]
\includegraphics[width=.55\textwidth]{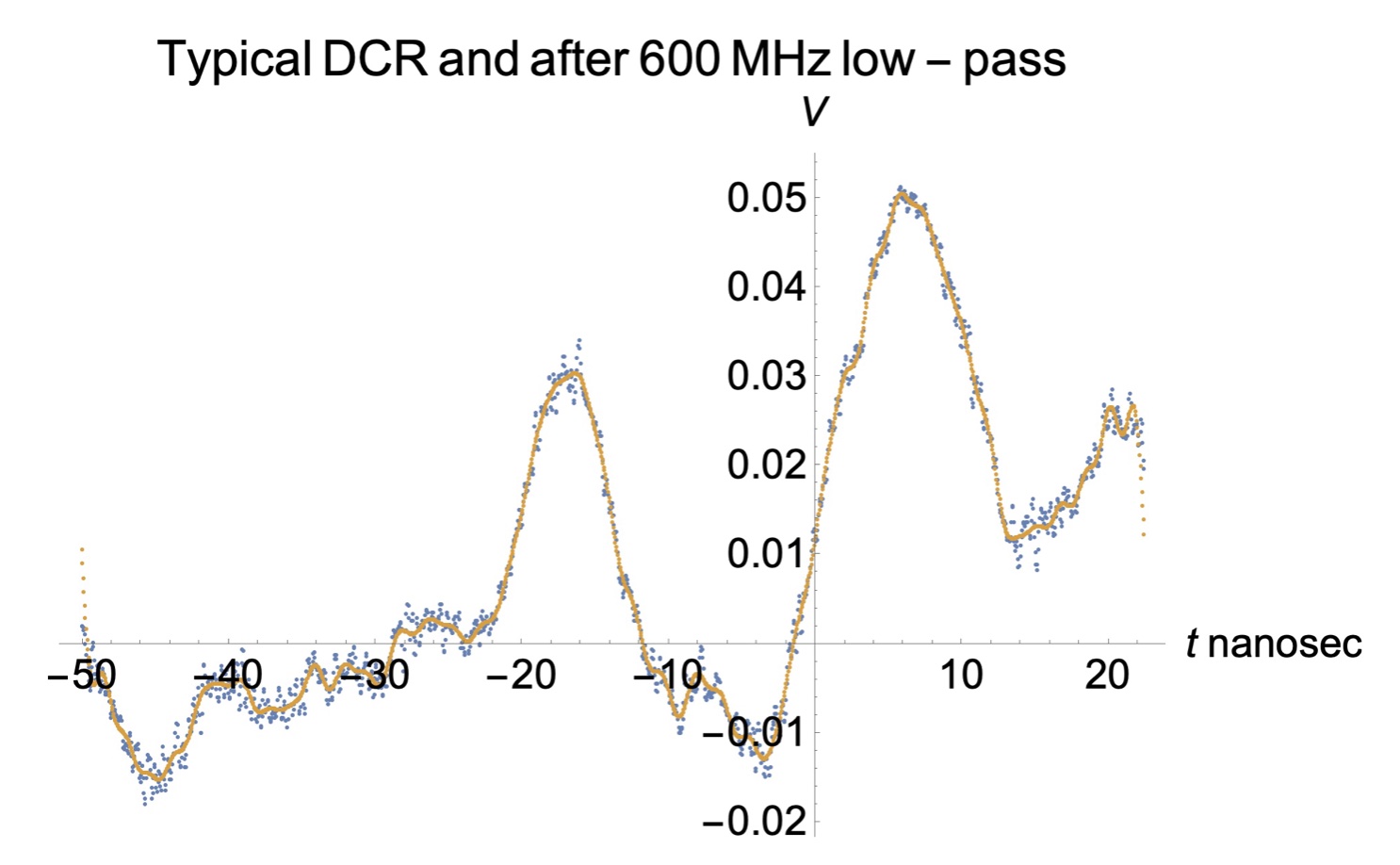}
\includegraphics[width=.4\textwidth]{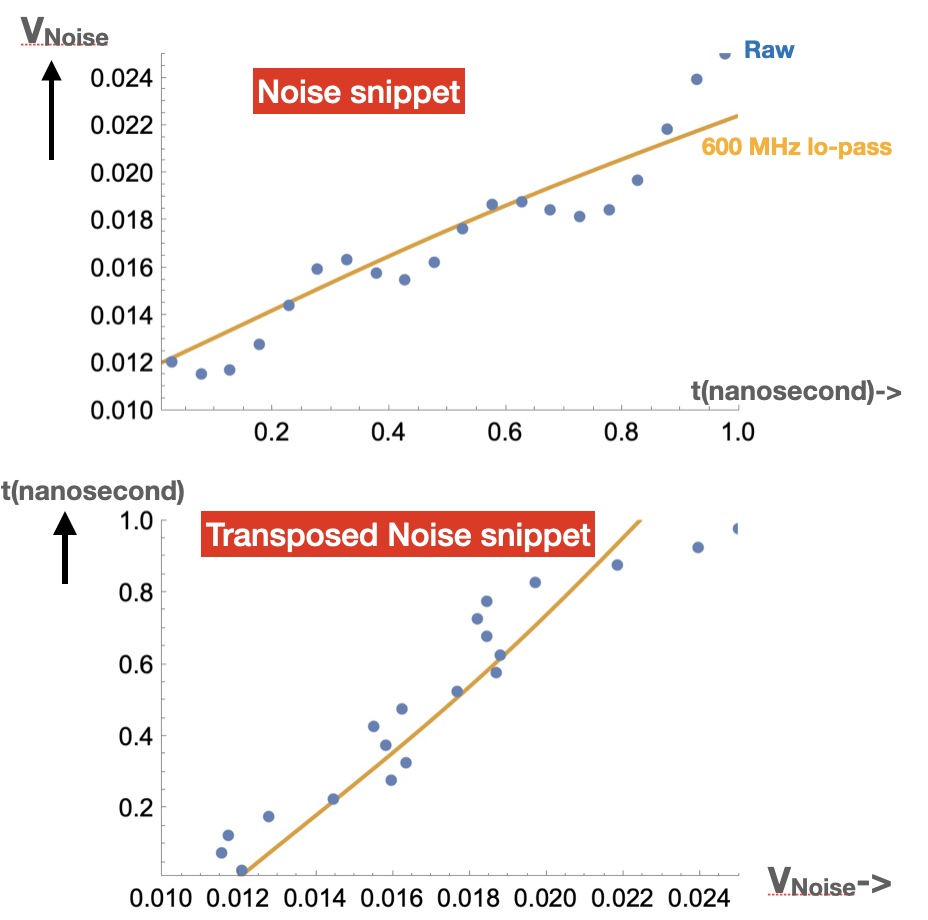}
\caption{left) A typical noise waveform showing full bandwidth data (blue points) and the filtered waveform (600 MHz cutoff - yellow curve). The seemingly innocuous high frequency content
can have a significant impact on timing jitter as illustrated in the right panel. This illustrates the importance of good design practice (ie bandwidth matching) to limit the high frequency noise. The latter could otherwise render ineffective the technique presented in this paper.}
\label{fig:imodel}       
\end{figure}

\subsubsection{ The Critical Role of Bandwidth Limit to Suppress High Frequency Noise}

In Figure 4 there is also evidence for RF pickup in these data. However, rather than trying to shield the sensor from this pickup source- presumably the local digital mobile network (GSM)- we chose, instead, to  record data at full bandwidth. This noise is, in any case, easily distinguishable from signal and is removable by offline digital signal processing (600 MHz low-pass filter).

As we will see below, turning on and off the high frequency noise proved to be very useful for understanding the robustness of the DCR jitter correction technique presented here.

But first we illustrate the importance of  eliminating high frequency noise, even when it appears to be a small component in the power spectrum. In Figure 5 we zoom in on a random trace of the
noise in our laser data. The left panel shows the scattered points before applying a 600 MHz low pass filter as well as a smooth trace once it is applied. In the right panel we further zoom in on 
the pulse and as a transpose (between time and amplitude) of a noise snippet.

Since the 1/f noise mitigation tool we have developed requires exploiting measurement of the noise contribution to slope and amplitude at a given threshold, Figure 5 illustrates the critical importance of bandwidth limiting.

We now perform a scan of the laser data set and find that the effectiveness of the mitigation tool is essentially short-circuited by the presence of high frequency noise. 
In Figure 6 we have followed essentially the same strategy as was illustrated in the model of Figure 3. We scan through six possible threshold settings and observe the time spread as
well as possible mitigation through  timeshift-Slope correlation. In the upper panel the time spread is relatively large and any such correlation has been smeared out by the high
frequency noise. In the lower panel, where 600 MHz low pass filtering was applied, there is clear sign of correlation and hence prospect for a mitigation tool.

\begin{figure}[!htp]

\includegraphics[width=.7\textwidth]{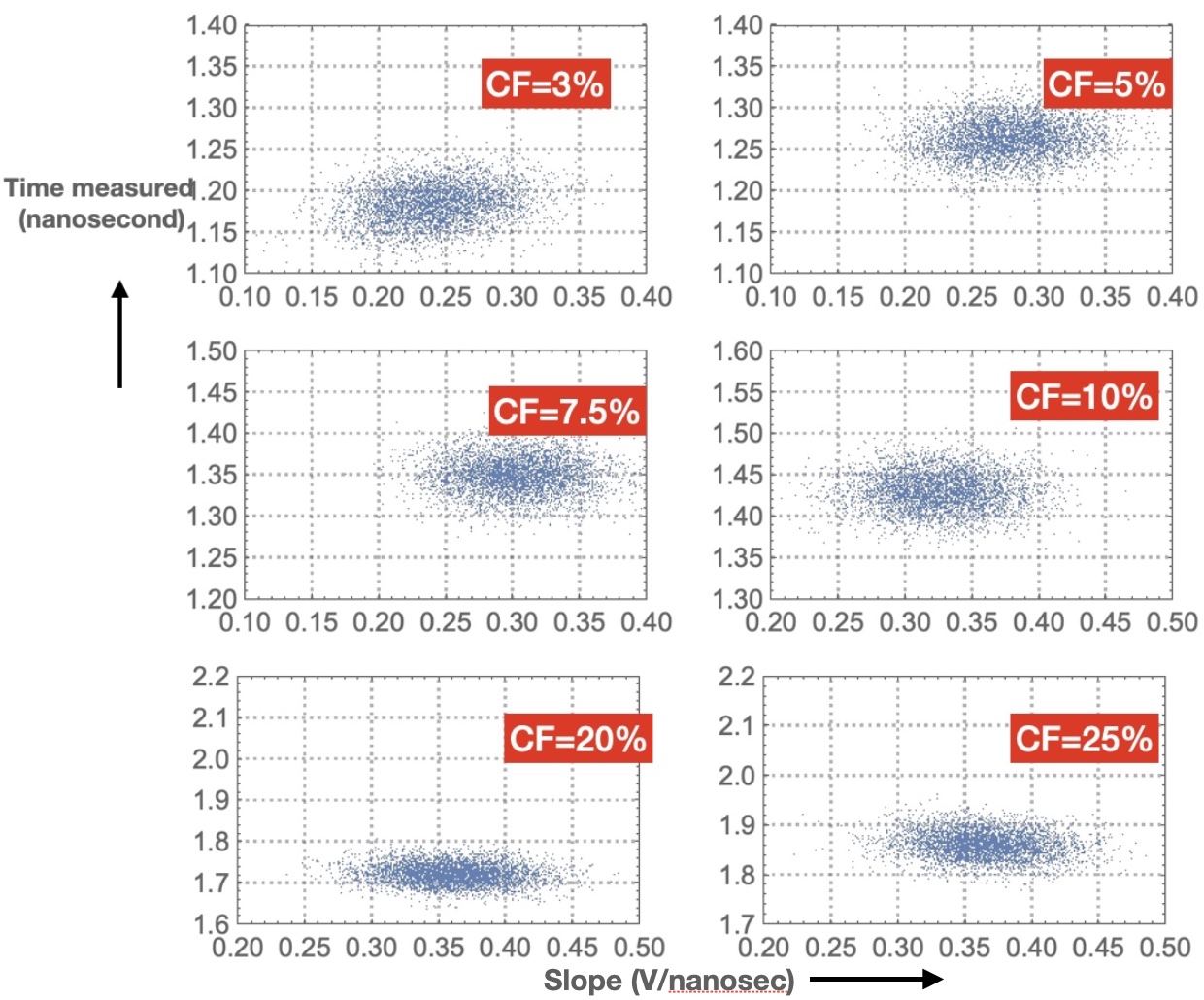}
\\
\includegraphics[width=.7\textwidth]{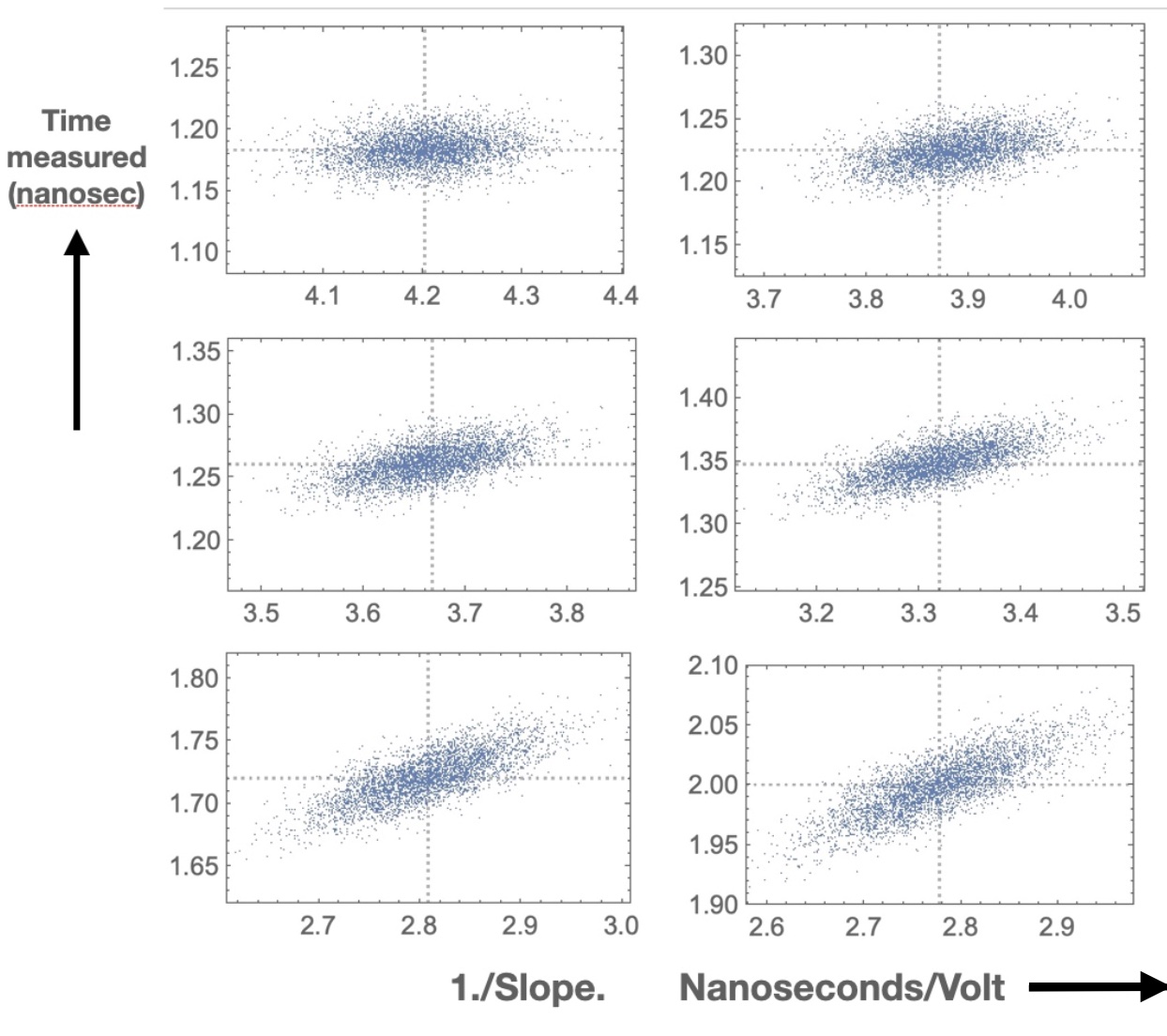}
\caption{Upper) Analogous analysis to that of Figure 3 (but replacing fixed frequency noise with the true spectrum) for several different Constant Fraction (CF) threshold settings before applying a low pass filter to suppress the RF pickup and... Lower) the same analysis after removing the RF pickup at high frequency. The x-coordinate is the inverse to that of Figure 3 since for the realistic case of fixed V$_{thresh}$, as with a discriminator, we are observing the inverse of the slope. Comparison of the two panels demonstrates the critical role of bandpass selection.}
\label{fig:imodel}       
\end{figure}

\subsection{Trading Active Baseline Subtraction with Single Delay Line Shaping}

In the following we repeat this analysis applying Single Delay Line (SDL) shaping as the method for baseline subtraction. We choose a delay of 0.6 nanoseconds, which happens to be a favored value 
for the CMS ASIC development \cite {tahereh}. The effect of this Delay Line Shaping can be seen in Figure 10, which illustrates the range of threshold settings that are appropriate for the
modified signal.

In principle the SDL technique could reduce the effectiveness of our mitigation tool by masking the time shift-Slope correlation. In practice we find, repeating the analysis with SDL shaped waveforms, that the mitigation tool remains effective as can be seen from Figure 7.

\begin{figure}
\centering
\centerline{\includegraphics[width=0.6\textwidth]{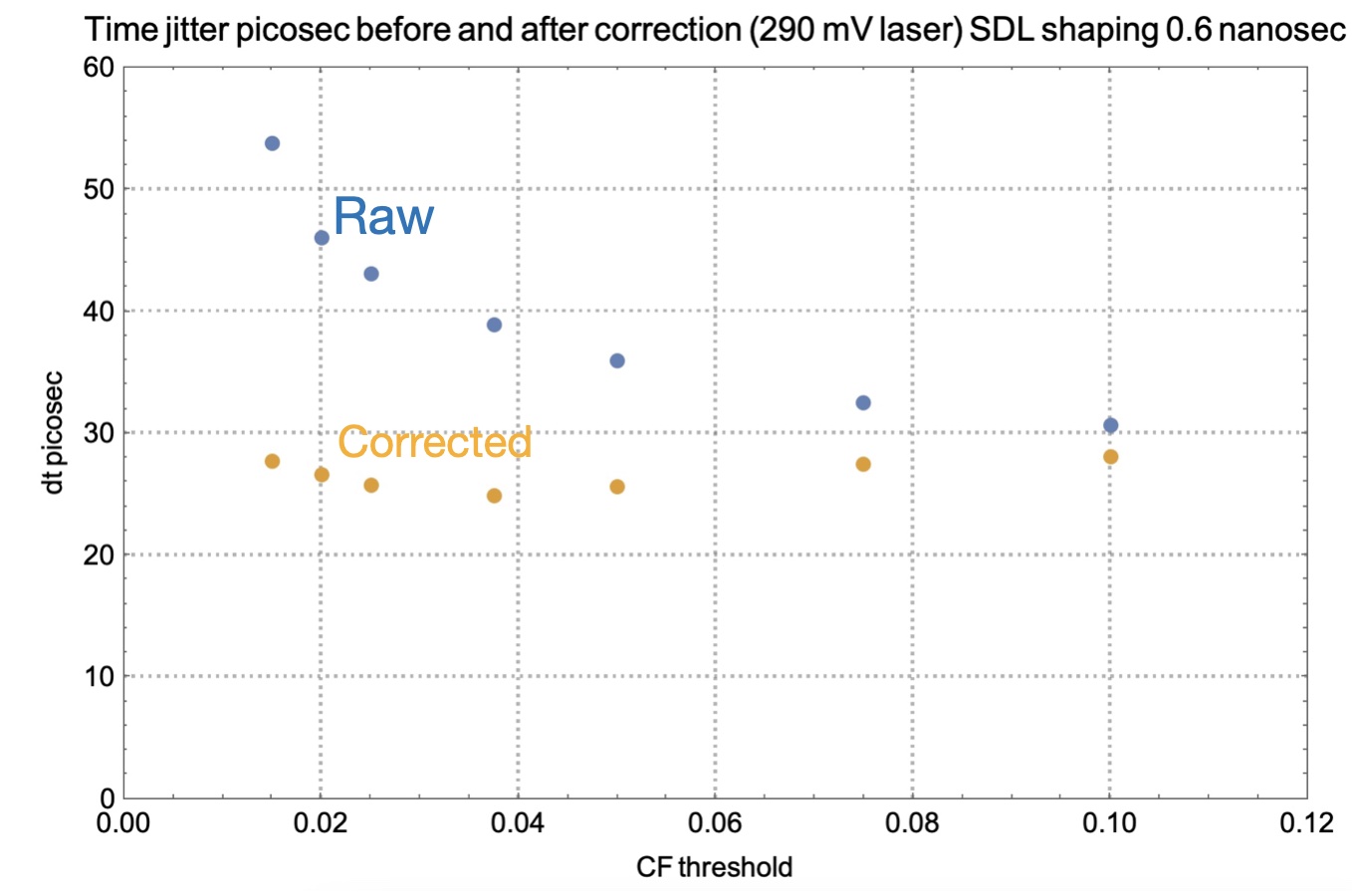}}
\caption{Repeating the analysis of the same data where we use Single Delay Line shaping (delay= 0.6 nanoseconds) instead of active baseline restoration on full waveforms we
once again see the gains from the same DCR mitigation tool. }
\label{fig:wavefor1}       
\end{figure}

\subsection{ Overall Result with Complete Laser Data Set}

Finally, we repeat the analysis using the actual data for both the laser signal and the noise. Since the laser data have $\sim\pm 8\%$ amplitude fluctuations, the data were first amplitude 
corrected to give the same charge integral for all pulses. This is the standard "Amplitude Walk Correction" applied as an initial pass to the data. From then on the analysis follows exactly the above procedure and the "DCR mitigator", as in Figure 8, is an empirical correction to times derived from timeshift -Slope correlation.

The tool is clearly effective with some dependence on the threshold choice. Depending on the application, this threshold dependence could be significant. For example, a mitigation tool that
favors lower threshold could have a larger impact in applications where photo-statistics govern the resolution (and low thresholds are favorable).

As laid out in the introduction we have developed the case for a tool which mitigates timing degradation due to low frequency noise. In all cases discussed here it appears to be effective. We should take note of the vulnerabilities- for example, to high frequency noise. Clearly it is important in any timing application to limit the effective bandwidth to best match the signal properties.

\begin{figure}
\centering
\centerline{\includegraphics[width=0.6\textwidth]{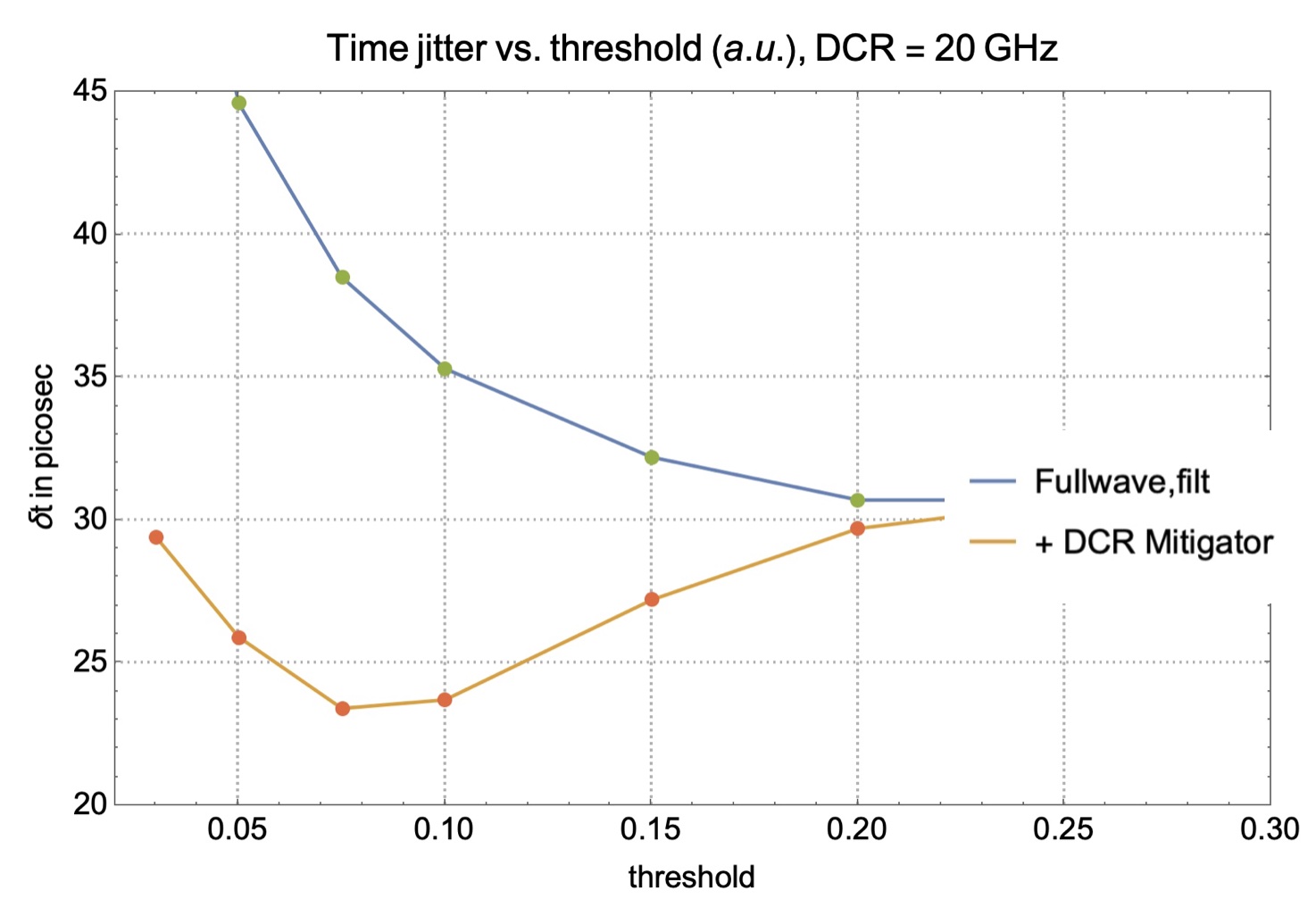}}
\caption{Time resolution for full analysis of laser data (i.e. actual laser signals with Constant Fraction Amplitude correction and actual noise). Note that these data were also processed with a 
600 MHz low pass filter, as discussed in the text. }
\label{fig:wavefor11}       
\end{figure}

\subsection{ Demonstration with a LYSO/SiPM prototype from the U. Virginia beam test}

We cannot leave this topic without demonstrating its applicability to an example with real radiation damaged SiPMs from a CMS prototype detector. The 3x3x57 mm$^3$ LYSO bar mentioned 
above with 3x3 mm$^2$ SiPMs glued to each end were exposed to a 120 GeV proton beam at FNAL \cite{UVatest}. For the plots shown in Figure 9 we see a very clear distinction between a bar
with un-irradiated SiPMs (and negligible Dark Count Rate) when compared to the one with previously irradiated SiPMs (resulting in a Dark Count Rate of $\sim13$ GHz). It is clear that, once again, the  timeshift -Slope correlation is in evidence and is a potentially useful tool.
Further discussion of the CMS detector case will be covered elsewhere since it is only relevant when applied to the particular ASIC used for that application.

\begin{figure}
\centering
\centerline{\includegraphics[width=0.85\textwidth]{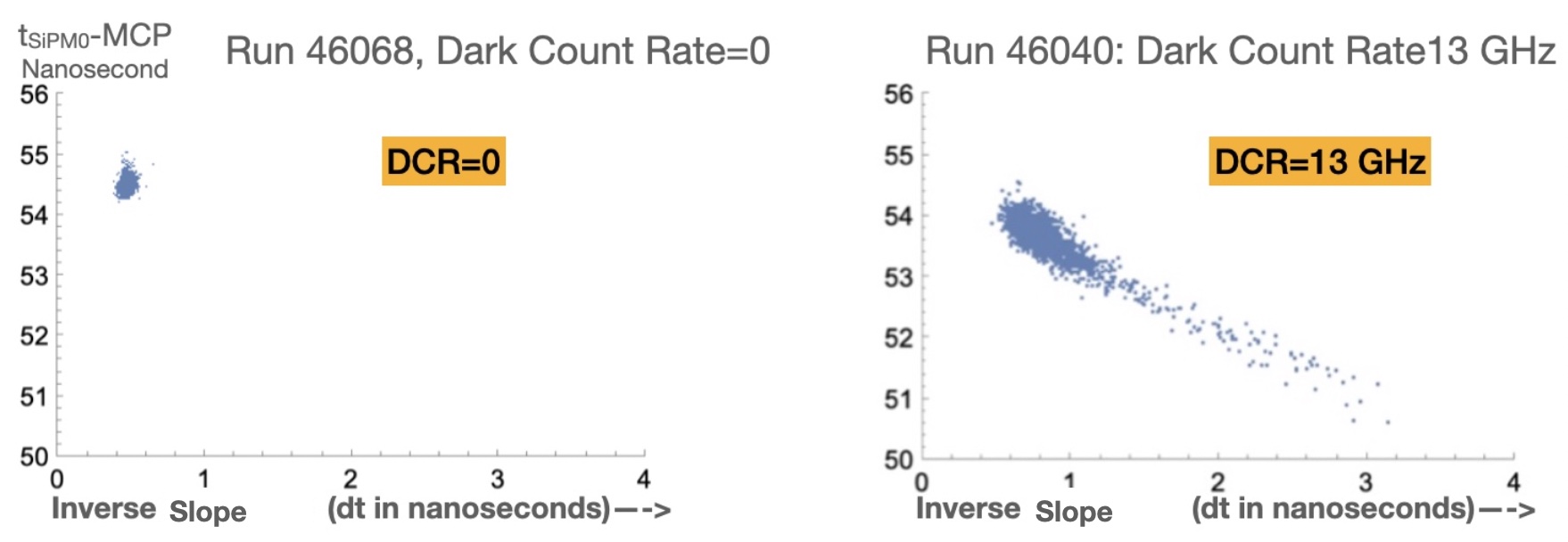}}
\caption{The same analysis was applied to testbeam data of a LYSO/SiPM model detector for the CMS Barrel Timing upgrade\cite{TDR} in a 120 GeV proton beam at FNAL. The correlation between slope and time shift is seen clearly using highly radiation damaged SiPMs with DCR=13 GHz (right panel) whereas, as expected, time spread and slope spread were insignificant when using undamaged SiPMs (left panel) }
\label{fig:slew}       
\end{figure}

\section{Conclusion and Acknowledgement}

	We have developed the case that in a timing detector system subject to low frequency noise with an, inevitable, baseline subtraction scheme an additional measurement- beyond signal amplitude and threshold crossing time, such as signal slope at threshold) can mitigate the timing degradation caused by this noise. 
	
	The domain of interest clearly falls somewhere between the usual discussion of high frequency noise in timing and the tool of baseline restoration/subtraction of low frequency noise employed in a variety of applications. 

	The case demonstrated in this report, of SiPM based timing in the presence of Dark Count Noise (which, after all, consists of randomly distributed SiPM responses to photoelectron equivalent input) focuses on an application where signal and noise have similar frequency spectra. This complements filtering or bandwidth matching techniques discussed for fast timing.

\begin{figure}[!htp]
\includegraphics[width=.45\textwidth]{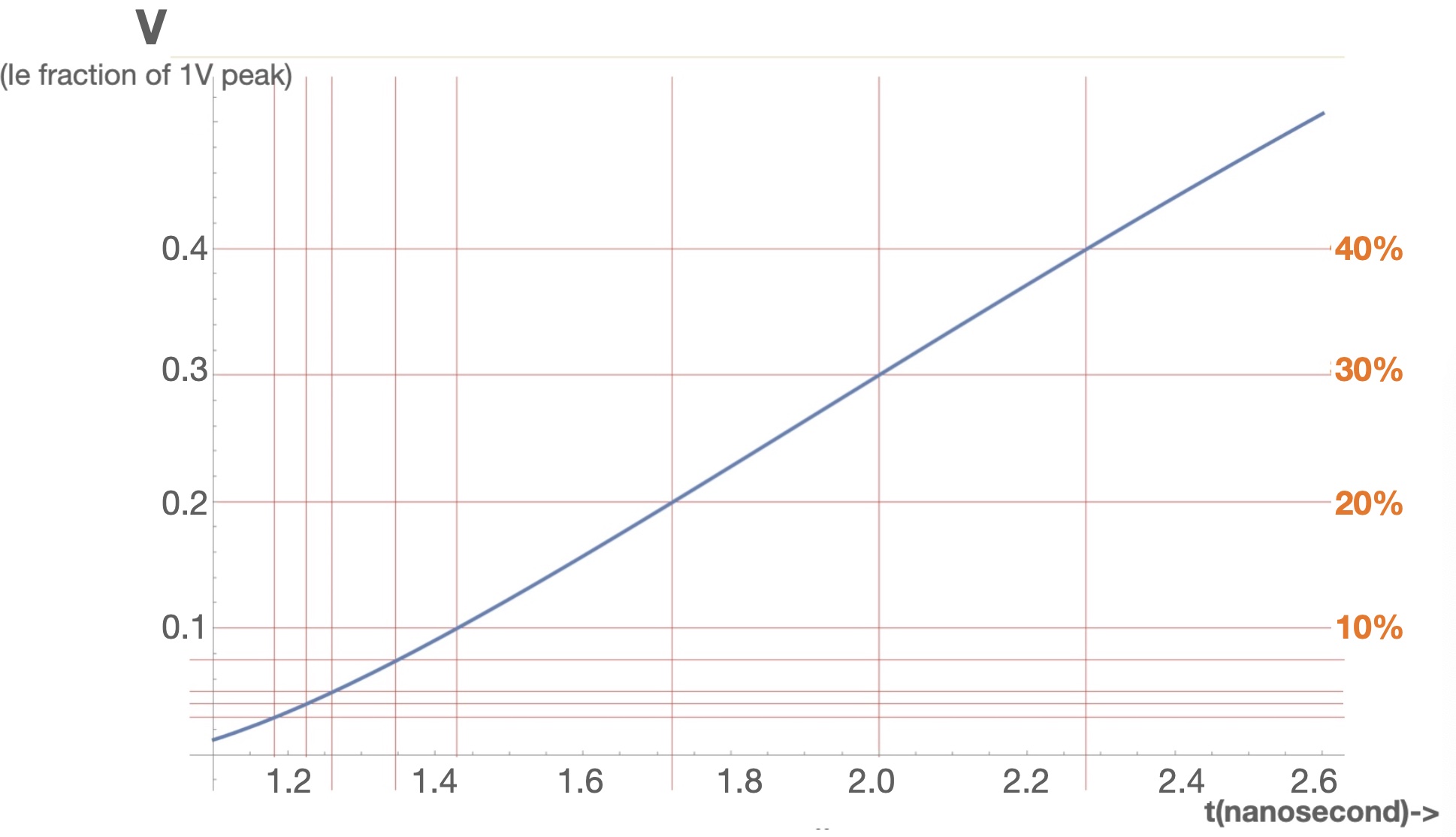}
\includegraphics[width=.55\textwidth]{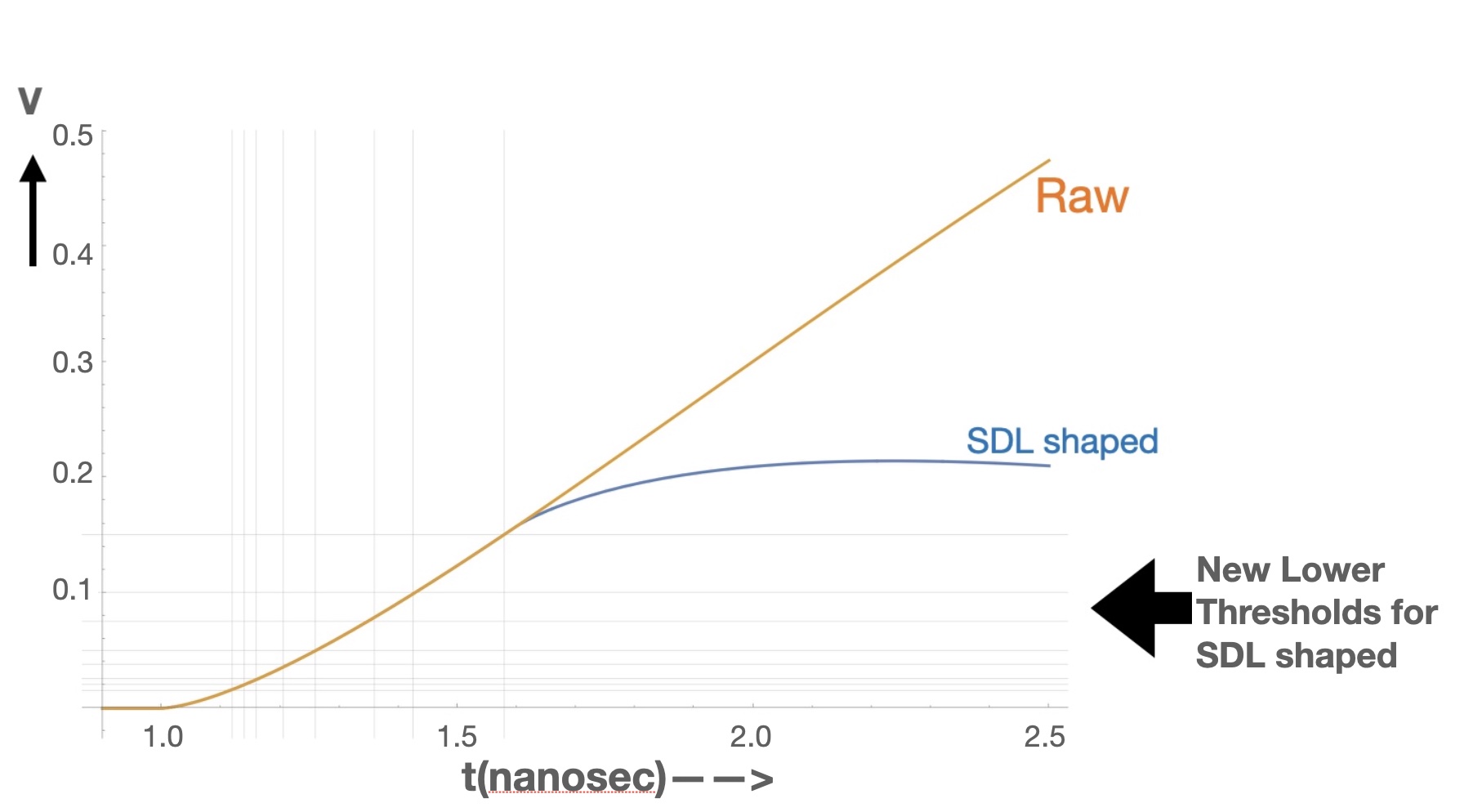}
\caption{For the remaining analysis of Model Signal + real Data we illustrate the series of threshold settings that are applied. For the 2nd part where we apply Single Delay Line Shaping
(with a delay of 0.6 nanoseconds) the peak amplitude is reduced by almost a factor of 5 so there is a corresponding decrease in thresholds(right) .}
\label{fig:imodel0}       
\end{figure}

	There is, of course, an extensive literature discussing this type of Power Spectrum (see, for example \cite{Radeka,Press,MOS}) but, so far as we know, a very limited number in the context of timing\footnote[4]{ An interesting example, however, can be found in \cite{Groth} where this feature occurs in time series from astronomical data.}. 
	
	We are simply left with the fact that both noise and signal have similar power spectra. Rather than turning to general filtering or weighting techniques \cite{Radeka} we are looking for a redundancy in the measurement from which to derive an ad hoc correction procedure.

\begin{figure}[!htp]
\includegraphics[width=.45\textwidth]{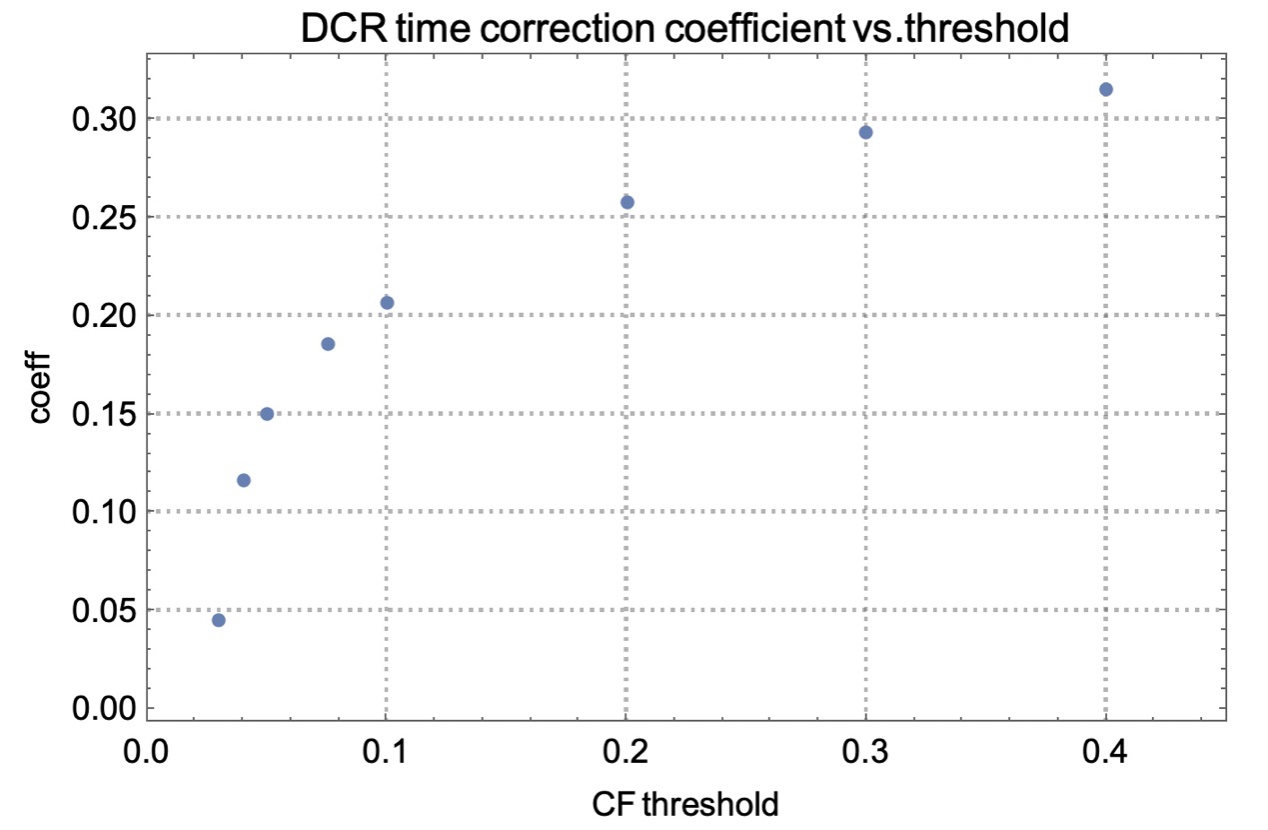}
\includegraphics[width=.45\textwidth]{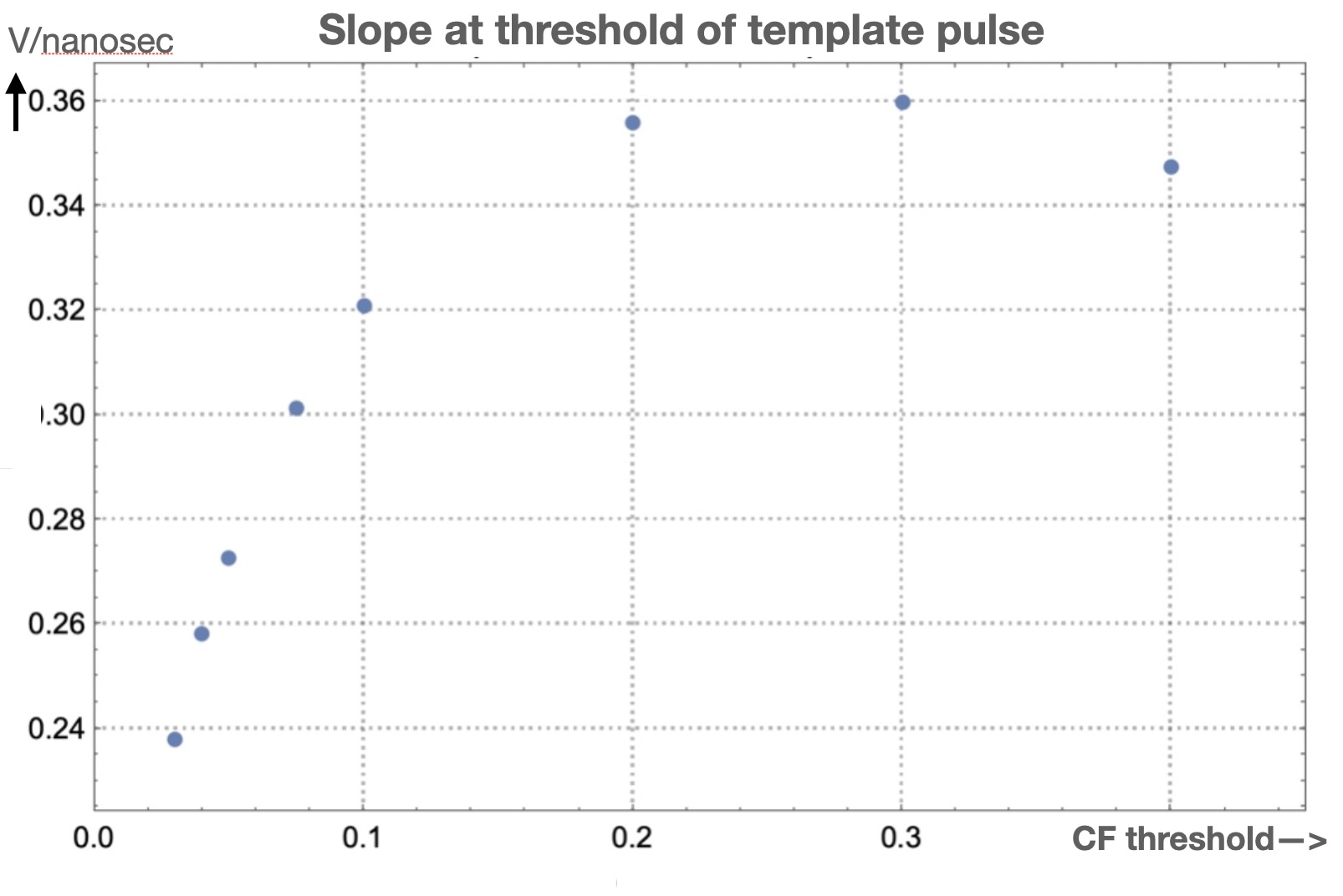}
\caption{As discussed in the text, we expect some similarity in the behavior of the correction coefficient (left panel) and the signal slope vs. threshold (right panel).}
\label{fig:imodel2}       
\end{figure}

	For the case of SiPM timing one might ask at what Dark Count Rate the tool developed here is likely to be effective. Given the time response shown in Figure 1 one might expect improvements even below DCR$\sim1$ GHz depending on the relative signal amplitude. We have here demonstrated that in our implementation (ie using complete waveform analysis) 
	the tool would be beneficial at DCR levels that will pertain to conditions at the CMS Barrel Timing Layer for most of the HL-LHC lifetime.
	
	Of course in such a large system it would be impractical to capture full waveforms at high sampling rate. However there is no reason that an ASIC incorporating Charge measurement and
	more than one timing threshold (such as the TOFHIR ASIC designed for the CMS BTL) couldn't achieve the subsidiary measurement of V'[t$_{thresh}$]. Today there are a number of
	ASICs which incorporate subsidiary measurements \cite{GianLuigi} amenable to the approach we have advocated.
	
		This work received partial support through the US CMS program under DOE contract No. DE-AC02-07CH11359.

\section { Appendix on Similarity between Correction Coefficients and signal shape.}

There are actually different meanings for the threshold settings used in the above discussion. That is because, as mentioned earlier, the Single Delay Line processed signals have a
much reduced peak amplitude (roughly $20\%$) as illustrated in Figure 10. For the following and Figure 11, we refer to Figure 10-left for the definition of threshold values.

As was remarked in section 2.2, the simple derivation at the beginning of this article relates the correction coefficients used in the DCR Mitigation tool to the signal shape. Not surprisingly
then, when we put these terms side-by-side in Figure 11, we clearly see the similarity.

\end{document}